\documentclass{mem}
\usepackage{natbib}\usepackage{txfonts}\usepackage{balance}
\usepackage{flushend}
\usepackage{graphicx}
\usepackage{xspace}
\usepackage{xcolor}
\usepackage[breaklinks,pdftex]{hyperref}
\usepackage{subfig}

\idline{75}{282}
\begin{document}

\newcommand{\dg}{$^{\circ}$\xspace}
\newcommand{\ecut}{$E_{\rm{cut}}$\xspace}
\newcommand{\dfu}{TeV$^{-1}$ cm$^{-2}$ s$^{-1}$\xspace}
\newcommand{\red}[1]{{\textcolor{red}{#1}}}

\title{
On the cosmic-ray distribution in the Galactic Center region: New insights from H.E.S.S.
}

   \subtitle{}

\author{
J. \, Devin\inst{1, 2} 
\and A. \, Lemière\inst{1} \and K. \, Streil\inst{3} \and R. \, Terrier\inst{1} \and C. \, van Eldik\inst{3} 
\\
on behalf of the H.E.S.S. collaboration
}

\institute{
Université de Paris, CNRS, Astroparticule et Cosmologie, F-75006 Paris, France
\and
Laboratoire Univers et Particules de Montpellier, Universit\'e de Montpellier, CNRS/IN2P3, F-34095 Montpellier, France, \email{devin@lupm.in2p3.fr}
\and 
Friedrich-Alexander-Universität Erlangen-Nürnberg, Erlangen Centre
for Astroparticle Physics, 91058 Erlangen,
Germany}

\authorrunning{H.E.S.S. Collaboration}

\titlerunning{The Galactic Center ridge emission}

\date{Received: ; Accepted: }

\abstract{
We performed a spectro-morphological analysis of the diffuse emission in the Galactic center region with H.E.S.S. $\gamma$-ray data. We relied on templates to model the diffuse emissions (the Galactic center ridge and the foreground component) and on a 3D likelihood fitting approach. We first assessed the validity of a continuous injection scenario near the Galactic center by investigating possible deviations from a $1/r$ profile of the cosmic-ray distribution and potential spectral variations within the Galactic center ridge. We found the data can appropriately be described by a scenario in which a steady source near the Galactic center continuously injects cosmic rays which diffuse through the Central Molecular Zone. We then derived the best-fit spectral parameters of the Galactic center ridge emission and we found a spectral transition near 10$-$20 TeV. 
}
\maketitle{}

\section{Introduction}

\begin{figure*}[ht!]
    \centering
    \includegraphics[scale=0.5]{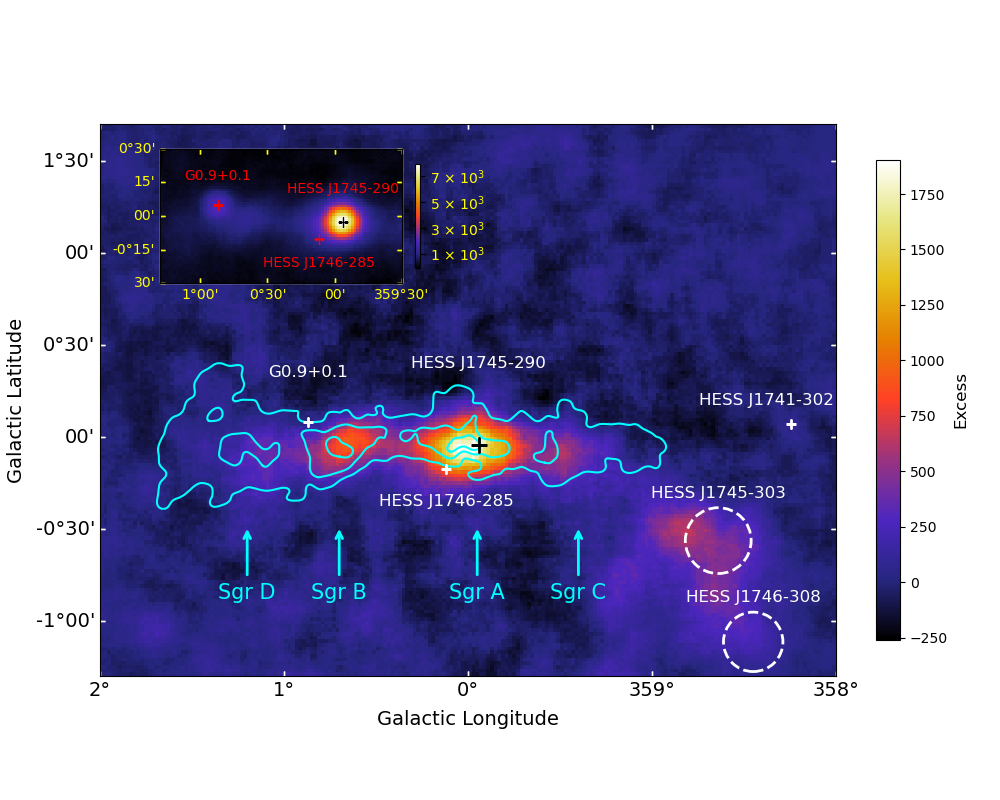}
    \caption{H.E.S.S. excess map from 400~GeV to 100~TeV after subtracting the contribution from HESS~J1745$-$290 and the PWN G0.9+0.1, showing the GC ridge emission (the initial excess map is given in the inset plot). The CS contours trace the CMZ and the white crosses and circles indicate the known H.E.S.S. sources in the region (with HESS J1745$-$290 in black).}
    \label{fig:excess_map}
\end{figure*}

The Galactic center (GC) is one of the richest regions in the Milky Way, harboring a large variety of potential particle accelerators such as supernova remnants, pulsar wind nebulae (PWNe) and young massive stellar clusters (the Arches, the Quintuplet and the Central cluster). The supermassive black hole Sgr A*, lying at the GC, is surrounded by dense molecular complexes that shape the so-called Central Molecular Zone (CMZ). Very-high-energy (VHE) $\gamma$-ray emission was reported towards this region that correlates with the gas content modulated by a non-uniform cosmic-ray (CR) distribution \citep{HESS_GC_ridge_discovery:2006, HESS_GC_PeVatron:2016, HESS_Diffuse_emission:2018, MAGIC_diffuse_GC:2020, VERITAS_GC:2021}. These studies confirmed an excess of CRs towards the GC and revealed a $1/r$ CR density profile indicative of a continuous injection and homogeneous diffusion through the CMZ. H.E.S.S. and VERITAS experiments did not report any significant spectral curvature in the GC ridge emission \citep{HESS_GC_PeVatron:2016, VERITAS_GC:2021} while MAGIC reported a possible curvature with a cutoff energy around 17 TeV at a 2$\sigma$-confidence level \citep{MAGIC_diffuse_GC:2020}. At higher energies, HAWC recently measured a spectral index of $\Gamma = 2.88 \pm 0.15$ \citep{HAWC:2024_GC}, significantly softer than the one reported by H.E.S.S. of $\Gamma = 2.32 \pm 0.05$ \citep{HESS_GC_PeVatron:2016}. We took advantage of a 3D likelihood analysis 
\citep[performed with \textit{Gammapy}\footnote{\url{https://docs.gammapy.org/1.2/}}, ][]{gammapy_paper:2023} which allows to fit simultaneously the position, the morphology and the spectrum of the different components, and used this method in revisiting the GC ridge emission with a larger H.E.S.S. dataset compared to previous studies. The increased statistics coupled with a better modeling of our main systematic uncertainties (related to the hadronic residual background and the foreground component) allow to better measure the GC ridge spectrum.

\section{H.E.S.S. data analysis}

We used 16 years of data reconstructed with the CT1$-$4 telescopes.  We selected observations with a maximum zenith angle of 40\dg (resulting in a mean zenith angle of 18\dg) and we applied a maximum photon direction offset to the camera center of 2\dg. Each calibrated run that passed the quality criteria was analyzed with a configuration optimized for Galactic sources within the H.E.S.S. analysis package framework described in \cite{Khelifi:2016}, including a Hillas-type shower reconstruction \citep{Hillas:1985} and a multi-variate analysis technique \citep{Becherini:2012} for the $\gamma$-hadron discrimination. The results were cross-checked using independent calibration, reconstruction \citep{Parsons:2014} and $\gamma$-hadron discrimination \citep{Ohm:2009} methods.

We performed an analysis in a 6\dg $\times$ 4\dg region, with a bin size of 0.02\dg and 30 (70) equally logarithmically-spaced energy bins from 200 (80) GeV to 100 (150) TeV for the reconstructed (true) energy, but we imposed a lower reconstructed energy cut of 400~GeV to limit the impact of the systematic uncertainties on the effective area at lower energies. For each observation, we normalized the background model on counts in OFF regions (where there is no $\gamma$-ray signal), including nuisance parameters in each energy bin. Observations were then stacked into one single dataset. For the analysis, we masked the crowded region of HESS~J1745$-$303 and HESS~J1746$-$308. Figure~\ref{fig:excess_map} depicts the residual excess map after taking into account the contribution from the two brightest sources (HESS~J1745$-$290 and the PWN G0.9+0.1, shown in the inset plot) and highlights the bright GC ridge emission spatially correlated with molecular material as traced by CS line emission.

The model of the region contains the four previously detected H.E.S.S. sources: HESS~J1745$-$290, the PWN G0.9+0.1, HESS~J1741$-$302 and HESS~J1746$-$285 \citep[also known as the arc source,][]{HESS_Diffuse_emission:2018}. To describe the emission from the CRs in the CMZ, we used the 3D description ($l$, $b$, $v$) of the CS gas \citep{Tsuboi:1999} that we distributed along the line of sight ($z$) following the 2D distribution ($l$, $z$) of the CO gas made by \cite{Sawada:2004}. This assumes that the CS and CO gas distributions are similar and do not vary with the latitude. We then multiplied the 3D ($l$, $b$, $z$) CS cube by a CR density profile (1/$r^{\alpha}$) and we projected the resulting 3D $\gamma$-ray emission into a 2D map ($l$, $b$). We also modeled the foreground contribution using a 2D Gaussian distribution \citep[as in ][]{HESS_Diffuse_emission:2018, HGPS:2018} for which we fitted its $\sigma$ extent in latitude. Finally, the model also contains the hadronic residual background for which nuisance parameters in each energy bin allow to properly model the related sources of uncertainties.

We first derived the best-fit position of the H.E.S.S. sources and the best-fit $\sigma$ extent of the foreground component ($\sigma = 0.32^{\circ} \pm 0.22^{\circ})$. Testing for different spectral shapes, we found that a curvature is needed in the foreground component so we used a logarithmic parabola shape throughout the analysis. The spectral shape of this foreground component is a source of systematic uncertainty and we checked that this choice does not change our conclusions.

\section{Morphology and Spectrum of the Galactic Center ridge}

\subsection{Validity of the steady-source scenario}

\begin{figure}
\resizebox{\hsize}{!}{\includegraphics[clip=true]{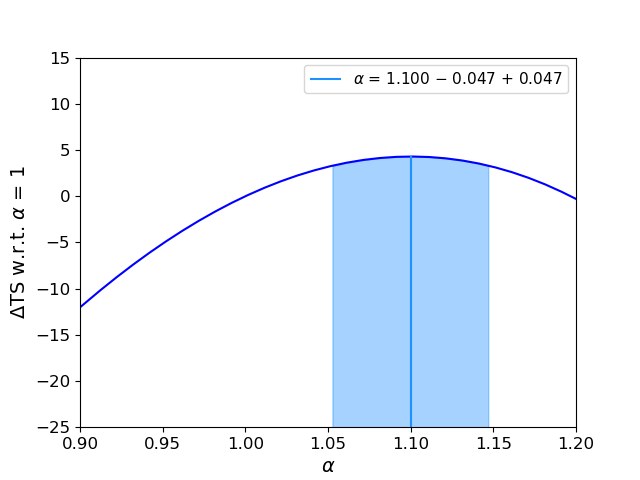}}
\caption{
\footnotesize
$\Delta$TS values obtained for different values of $\alpha$ (with respect to $\alpha = 1$) using a 3D $(l, b, z)$ distribution of the CS gas in the CMZ.}
\label{fig:Fit_alpha}
\end{figure}

We first assessed the validity of a continuous injection model, which implies a $1/r$ CR density profile and a constant spectral index within the GC ridge. Using our best-fit model, we performed a fit with the gas template multiplied by a $1/r^{\alpha}$ profile for different values of $\alpha$. The spectral parameters of all the components were let free during the fit. The obtained Test Statistic (TS, being twice the likelihood difference) profile (with respect to $\alpha = 1$) is given in Figure~\ref{fig:Fit_alpha} and indicates a best-fit value of $\alpha = 1.10_{-0.05}^{+0.05}$. This is consistent with the value derived by \cite{MAGIC_diffuse_GC:2020} of $\alpha = 1.2 \pm 0.3$ and by \cite{HESS_GC_PeVatron:2016} of $\alpha = 1.10 \pm 0.12$. This result indicates that the morphology is compatible with a continuous injection near the GC, given the gas distribution we used. To assess any possible spectral variations across the GC ridge, we used the 2D velocity-integrated CS map that we spatially divided into 7 regions (containing for example the molecular complexes Sgr B2, Sgr C, MC 20 km s$^{-1}$ and MC 50 km s$^{-1}$ clouds). We described their emission by a power-law model and we fit their spectral parameters simultaneously with those of the nearby H.E.S.S. sources and of the foreground and background components. We found no significant spectral variations across the GC ridge, strengthening the validity of the steady-source scenario.

\subsection{Gamma-ray spectrum}

\begin{figure*}[ht!]
    \centering
    \subfloat{{\includegraphics[width=6cm]{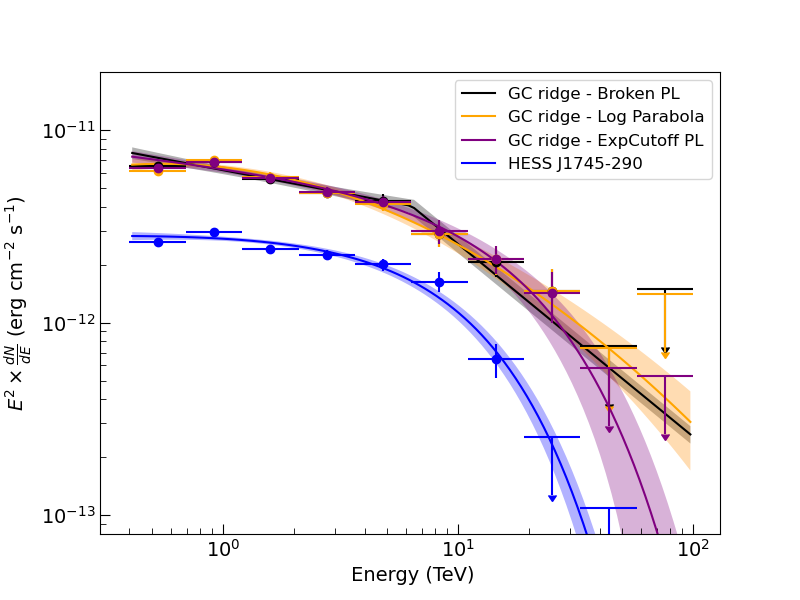} }}%
    \qquad
    \subfloat{{\includegraphics[width=6.5cm]{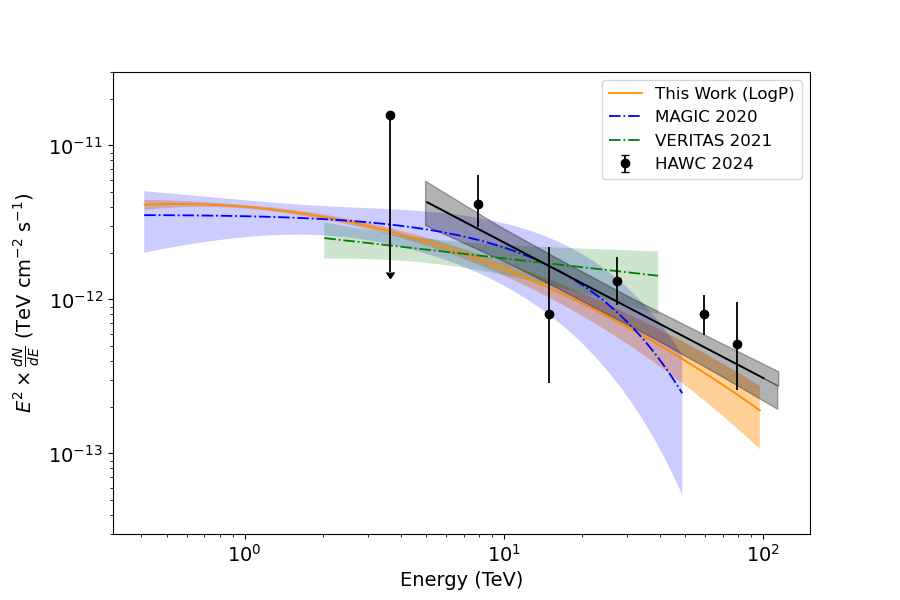} }}%
    \caption{(Left) Best-fit spectral models (and corresponding spectral energy distribution) of the GC ridge using different spectral shapes. (Right) Comparison with other experiments (references can be found in the text).}
    \label{fig:spec_ridge}
\end{figure*}

Using a $1/r$ CR density profile (fixing $\alpha = 1$), we fit the emission of the GC ridge with different spectral models: a power law (PL), a broken power law (BPL), a logarithmic parabola (LogP) and a power law with an exponential cutoff (ECPL), leaving the parameters of all the components free. The results are given in Table~\ref{tab:CMZ_spec} with the associated statistical and systematic uncertainties. The systematic uncertainties were evaluated through Monte-Carlo simulations, introducing a mismodeling on the instrument response functions (acceptance and energy bias) and on the residual hadronic background spectral model. The $\Delta$TS values given in Table~\ref{tab:CMZ_spec} are calculated with the PL as a reference model, and show that a curved spectrum is preferred at a 3$\sigma$ confidence level. Given these $\Delta$TS values, we are not able to distinguish between the different curved models tested (BPL, LogP or ECPL) within the H.E.S.S. energy range. Figure~\ref{fig:spec_ridge} (left) shows that these best-fit spectral models (and corresponding spectral energy distribution) are indeed similar up to $\sim$ 40 TeV.

\begin{table*}[t!]
    \centering
    \scriptsize
    \begin{tabular}{l|cccc|c}
    	\hline
        \hline
        Model& $N_0$ (\dfu) & $\Gamma$, $\Gamma_1$ or $\alpha$  & $\Gamma_2$ or $\beta$ & $E_{\rm{b}}$ or \ecut (TeV) &   $\Delta$TS \\
		\hline
        PL & $(3.89 \pm 0.15 \pm 0.92_{\rm{syst}}) \times 10^{-12}$ & $2.35 \pm 0.04 \pm 0.10_{\rm{syst}}$ & $-$ & $-$ & 0 \\
        BPL & $(3.96 \pm 0.15 \pm 0.44_{\rm{syst}}) \times 10^{-12}$ & $2.24 \pm 0.05 \pm 0.10_{\rm{syst}}$ & 2.88 (fixed) & $6.30 \pm 0.22 \pm 0.72_{\rm{syst}}$ & 11.5 \\
        LogP & $(4.00 \pm 0.10 \pm 0.76_{\rm{syst}}) \times 10^{-12}$ & $2.14 \pm 0.06 \pm 0.11_{\rm{syst}}$ & $0.12 \pm 0.03 \pm 0.05_{\rm{syst}}$ & $-$ & 14.1\\
        ECPL & $(4.11 \pm 0.18 \pm 0.93_{\rm{syst}}) \times 10^{-12}$ & $2.14 \pm 0.06 \pm 0.10_{\rm{syst}}$ & $-$ & $18.42 \pm 6.08 \pm 3.20_{\rm{syst}}$ & 14.5 \\
     \hline
    \end{tabular}
    \vspace{0.05cm}
    \caption{Best-fit spectral parameters for the Galactic Center ridge $\gamma$-ray emission. ECPL stands for power law (PL) with exponential cutoff: $dN/dE$ = $N_0$ ($E$/$E_0$)$^{-\Gamma}$exp($-E/E_{\rm{cut}}$), LogP for logarithmic parabola: $dN/dE$ = $N_0$ ($E / E_0$)$^{-\alpha - \beta \log(E/E_0)}$ and BPL for broken power law: $dN/dE$ = $N_0$ ($E/E_{\rm{b}}$)$^{-\Gamma}$ with $\Gamma$ = $\Gamma_1$ and $\Gamma_2$ before and after the energy break $E_{\rm{b}}$ respectively. The $\Delta$TS value is given with respect to the PL model and the flux normalization $N_0$ is evaluated at $E_0$ = 1 TeV for each model.}
    \label{tab:CMZ_spec}
\end{table*}

The amount of $\gamma$-ray emission associated with the GC ridge is compatible with previous H.E.S.S. publications \citep{HESS_GC_PeVatron:2016, HESS_Diffuse_emission:2018} and we now detect a spectral curvature. This is mainly thanks to an increased statistics and a better modeling of the main sources of systematic uncertainties when measuring such extended emission (related to the background and foreground components). The best-fit spectra we obtained are also consistent with those from MAGIC, VERITAS and HAWC, as illustrated in Figure~\ref{fig:spec_ridge} (right), showing that a significant spectral transition occurs near 10$-$20 TeV.

\section{Conclusion}

We revisited the diffuse emission in the GC region with H.E.S.S. data and a 3D likelihood approach ($l$, $b$, $E$). We relied on templates to describe the region with a self-consistent model, for which we simultaneously fit the spatial and spectral parameters of the different components. We also used a 3D CS gas and CR distributions ($l$, $b$, $z$) to properly model the GC ridge emission. We confirmed that the morphology and spectrum of the ridge are compatible with a steady-source injection model: the CR gradient is appropriately described by $1/r^{\alpha}$ with $\alpha = 1.10_{-0.05}^{+0.05}$ and no significant spectral variations are detected. Using $\alpha = 1$, we derived the best-fit spectral parameters of the GC ridge emission using different spectral shapes. We detected a significant curvature, with a spectral transition near 10$-$20 TeV. Theses results are consistent with those from other experiments such as MAGIC and HAWC. A better angular resolution in the HAWC energy range would be needed to properly understand the highest-energy part of the spectrum. This will be achieved with the next generation of Cherenkov telescopes CTAO, which will also better constrain the energy-dependent morphology and therefore the validity of the steady-source scenario.

\bibliographystyle{aa}
\bibliography{devin_main}

\end{document}